\documentclass[runningheads]{ola}

\usepackage{epsfig}
\usepackage{cite}
\usepackage{amsmath}
\usepackage{listings}
\usepackage{comment}
\usepackage{graphicx}
\usepackage{float}
\usepackage{hhline}

\usepackage{array}   %% habitual addition
\newcolumntype{C}{>{\centering\arraybackslash}X} 
\usepackage[english]{babel} % To obtain English text with the blindtext package
\usepackage{blindtext}

\def\url#1{\expandafter\string\csname #1\endcsname}

\begin{document}

\pagestyle{headings}

\mainmatter
% -------------------------
% Title and page formatting
% -------------------------

\title{Cluster images with AntClust: a clustering
algorithm based on the chemical recognition system
of ants}

% Title
\titlerunning{Cluster images with AntClust}

% Title for odd pages
\author{Winfried Gero Oed, and Parisa Memarmoshrefi}

% Authors for the top of the even pages
\authorrunning{W. Oed, and P. Memarmoshrefi}

\institute{
Department of Computer Science \\
Georg August University of Goettingen, Germany \\
\email{winfired.oed@stud.uni-goettingen.de}\\
\email{memarmoshrefi@informatik.uni-goettingen.de}\\
}

\maketitle

%---------------------------------------------------------
%               Text
%---------------------------------------------------------

%--------------------------------
%        Abstract
%--------------------------------
\begin{abstract}
We implement \textit{AntClust}, a clustering algorithm based on the chemical recognition system of ants and use it to cluster images of cars.
We will give a short recap summary of the main working principles of the algorithm as devised by the original paper \cite{b0}.
Further, we will describe how to define a similarity function for images and how the implementation is used to cluster images of cars from the vehicle re-identification data set.
We then test the clustering performance of \textit{AntClust} against DBSCAN, HDBSCAN and OPTICS.
Finally one of the core parts in \textit{AntClust}, the rule set can be easily redefined with our implementation, enabling a way for other bio-inspired algorithms to find rules in an automated process.
The implementation can be found on GitLab \cite{b8}.
\end{abstract}

%
% ----------------------------------
%  Introduction
% ----------------------------------
\section{Introduction}
In this work we describe the principle and usage of \textit{AntClust}, a clustering algorithm based on the chemical recognition system of ants, which was developed by Nicolas Labroche, Nicolas Monmarch´e and Gilles Venturi \cite{b0}.

The clustering algorithm's general idea is based on ant's chemical recognition system.
Ants form colonies where many individuals help the colony to survive by maintaining it.
To prevent harm from intruders the individual ants of a colony have developed a mechanism to recognize their nestmates.
Labroche et al. refer to the sources \cite{b2, b3} for more information.
Generally the nestmate recognition is based on odors.
These are individual to every ant, meaning that every ant has its own odor.
Additionally, every ant maintains an odor template which is generated based on its encounters with other ants.
If two ants meet, they will recognize each other's odors and if the odors are similar enough - the ants "like" to smell each other - they will accept, know each of them belongs to the same colony.
Rejection can happen if the odors do not match the template.
As the odor, the template is individual to every ant.
It is constantly updated during the encountering of other ants.
Thus there is no global colony template and the whole identification process is decentralized.
It is only defined in the sum of the different templates carried within every ant and is a dynamic process, subject to constant change.

The algorithmic idea from Labroche et al. is to create artificial ants which will be initialized with a certain genetic.
This artificial genetic is a data tuple of a data set, e.g., an image.
Based on this genetic, the ant will define its own template for recognizing potential nestmates and repel intruders.
This will generate artificial ant colonies or differently framed, it will form clusters.
We implement the described algorithm using Python and use our implementation to cluster images of cars, taken by public surveillance cameras.
In this clustering task, the different images of a distinct car form one particular cluster.
Images are taken from the \textit{vehicle re-identification (VeRi)} data set\cite{b6}.
Clustering images is not an easy task since clustering needs a similarity measure between its data tuples.
For this we construct a concept for a similarity function that work on images and use it in our implementation.

%

% \vspace{1em}
This paper is structured as follows.
First, there will be a general description of the algorithm - which is a summary of the original paper - in section \ref{sec:antclust_working_principle}.
In the following section \ref{sec:proposed_method}, we will describe our approach to construct a similarity function that works on images.
So then, the obtained results and the clustering performance will be shown in section \ref{sec:antclust_results}.
Finally, there is a general discussion about the algorithm, its implementation and possible improvements as well as thoughts on how to improve/combine it with other bio-inspired and evolutionary-based approaches in section \ref{sec:antclust_conclusion_future}.

% ----------------------------------
%  AntClust Theoretical Framework
% ----------------------------------
\section{AntClust Theoretical Framework}
\label{sec:antclust_working_principle}
This section is a summary of the original \textit{AntClust} paper \cite{b0}.
It explains how \textit{AntClust} works and how the algorithm is structured.
The AntClust algorithm can be summarized into the sub actions in table \ref{tab:antclust_steps}.
%\vspace{0em}
\begin{table}[ht]
    \centering
    \normalsize
    \caption{AntClust algorithmic steps}
    \begin{tabular}{ l l }
     Phase & \\
     \hline
     1 & initialize ants \\ 
     2 & initialize ant templates \\
     3 & randomly meet ants and apply meeting rules\\  
     4 & shrink ant colonies \\
     5 & re-assign ants with no colony
     \label{tab:ant_clust_steps}
    \end{tabular}
    \label{tab:antclust_steps}
\end{table}

These sub actions will now be described one after another.

For this section, it will be assumed that the data set to perform the clustering algorithm is a very simplistic one, having just one feature. This feature is just a real number between $0$ and $1$.
Therefore one tuple from the set would be just a number.
This simplifies in order to understand the working principle of \textit{AntClust}.

Phase one is the initialisation of the ants.
To represent the data set as an ant colony \textit{AntClust} will create artificial ants where each ant will correspond to one distinct data tuple from the data set -meaning in our example, the data set [1,2,3] will be represented by three ants, each holds one number as their genetics-.
Inside the computer these ants are program objects, an instance of a certain class.
Every ant has their own attributes.
% \vspace{1em}
%
%
% Ant attributes
\begin{itemize}
    \item[$\cdot$] Genetic of an ant is defined as one data tuple from the data set. 
    \item[$\cdot$] Label of an ant indicates to which colony or cluster the ant belongs.
    \item[$\cdot$]Template is distinct to every ant. The template is a real number $0 \leq x \leq 1$ and is used to estimate whether another ant is accepted during a meeting or not. The template is dynamic and will evolve over time.
    \item[$\cdot$]Age is an indicator of how many meetings the ant has had with other ants during the meeting phase of the algorithm.
    \item[$\cdot$]M is an estimator, $0 \leq M \leq 1$, that reflects how successful the ant is during its meetings with other ants. If the ant is not very well accepted during its meetings, M will decrease
    \item[$\cdot$]$M^+$ is an estimator, $0 \leq M^+ \leq 1$, which reflects how well the ant is accepted inside its nest / colony. This estimator is therefore updated whenever the ant is meeting with another ant and is reset to zero if the ant loses its label.
    \item[$\cdot$]$Max(Sim(ant, \cdot))$ is the maximal similarity obtained during a meeting with another ant. Here "$\cdot$" means all other ants an ant has met.
    \item[$\cdot$]$\overline{Sim}(ant, \cdot)$ is the mean similarity between this ant and all the ants this ant has met.
\end{itemize}
\vspace{3em}

To initialize an ant and finish phase $1$, the parameters will be set as follows:
\vspace{-1em}
\begin{itemize}
    \item[$\cdot$] Genetic $\longleftarrow i^{th}$ tuple of the data set.
    \item[$\cdot$] Label $\longleftarrow 0$, Age $\longleftarrow 0$, M $\longleftarrow 0$, $M^+$ $\longleftarrow 0$.
    \item[$\cdot$] $Max(Sim(ant, \cdot))$ $\longleftarrow 0$, $\overline{Sim}(ant, \cdot)$ $\longleftarrow 0$.
    \item[$\cdot$] Template $\longleftarrow 0$, will be initialized directly in the next step.
\end{itemize}

% \vspace{1em}
Initialisation of the template is done in phase $2$, where the template of an ant is defined by
\begin{equation*}
    \label{eq:ant_template}
    template \longleftarrow \frac{\overline{Sim}(ant, \cdot) + Max(Sim(ant, \cdot))}{2}
\end{equation*}

The template serves the function of determining a distance in by which the ants should accept each other or not.
It is a reference that estimates how far ants can be - in a distanced manner - away from each other but still belonging to the same colony.
To initialize it, the ant will have meetings with randomly selected ants.
The number of meetings can be freely defined, but Labroche et al. suggest that it is calculated \footnote{The iteration amount can be computed
by $0.5\cdot \alpha \cdot N$ , where N is
the number of ants, i.e. data tuples inside the data set. $\alpha=150$
was tested by Labroche et al. and found as the most general optimal
value. For some data sets better results can be obtained by tuning the parameter.}.
After these meetings, the ants will have formed their template and the algorithm can now continue with phase $3$.

% \vspace{1em}
In phase $3$, two ants are randomly chosen and meet each other.
Meeting in the sense of \textit{AntClust} means that a rule set is applied.
The rule set given by Labroche et al. is defined below.
Given rules will be applied one after another.
If no rule matches, then nothing will happen.
It is assumed that there is an $ant_i$ and an $ant_j$ with their corresponding variables $x_{i, j}$.
% \vspace{1em}

\begin{itemize}
    \item[$\cdot$] New colony creation rule (R1): \\
    $If (label_i = label_j) \; and \; Acceptance(i,j)$ then a new colony will be created by defining a new label $label_{new}$ and this label will be assigned to the ants\\
    $label_{i,j} \leftarrow label_{new}$.\\
    If the ants do not accept  each other, R6 is applied.
    \vspace{1em}
    
    \item[$\cdot$] Ant with no label is assigned to existing colony (R2):\\
    $If(label_i = 0 \land label_j \ne 0) \; and \; Acceptance(i,j)$ \\
    then $label_i \longleftarrow label_j$. This rule applies symmetrically if the condition is reversed.
    \vspace{1em}
    
    \item[$\cdot$] accepting of colony mates (R3): \\
    $If(label_i = label_j) \land (label_i \ne 0) \land (label_j \ne 0) \; and \; Acceptance(i,j)$ \\
    then increase $M_i, M_j, M_i^+, M_j^+$. \\
    Increasing means $x \leftarrow (1-\alpha)\cdot x + \alpha$, where $\alpha = 0.2$.
    \vspace{1em}
    
    \item[$\cdot$] not accepting of two colony mates (R4): \\
    $if(label_i = label_j) \land (label_i \ne 0) \land (label_j \ne 0) \; and \; Acceptance(i,j) = False$, increase $M_i, M_j$, decrease $M_i^+ , M_j^+$. \\
    Decreasing means $x \leftarrow (1-\alpha)\cdot x$, where $\alpha = 0.2$.
    Additionally, the ant with the smaller colony integration value - which is $ant_i$ $if(M_i^+ < M_j^+)$ or $ant_j$ if the condition is reversed - loses its label $(label \leftarrow 0)$ and does not belong to a colony anymore.
    \vspace{1em}
    
    \item[$\cdot$] meeting of colony different ants (R5): \\
    $if(label_i \neq label_j) \land Acceptance(i,j)$ decrease the colony size estimator $M_i , M_j$. The ant with the lower estimator will change its label and belongs to the colony of the other ant.
    \vspace{1em}
    
    \item[$\cdot$] Default rule (R6): \\
    If none of the above rules apply, nothing will happen during the meeting of the two ants.
    \vspace{1em}
    
\end{itemize}

The above rules give a working framework to form clusters as explained now.
In the very beginning  all ants are initialized with no label - i.e. $label \leftarrow 0$.
Thus when two ants meet in the beginning, mostly R1 will be applied.
This will create a lot of little clusters containing only two ants.
Once more and more ants have got a label assigned through  R1, it becomes more likely that R2 is applied and thus, the initial formed clusters begin to grow.
If two ants belong to the same cluster, there are two cases: they accept each other or they do not.
In the first case, R3 will be applied which will increase the cluster size estimator $M$ and the colony integrity estimator  $M^+$, which makes sense since it seems that the colony is quite big if two randomly chosen ants belong to the same colony, accepting each other means that the colony is still in a good integrity state.
In the second case, R4 will be applied as the colony mates do not accept each other.
This will increase the colony size estimator since the colony must be relatively big if two randomly chosen ants belong to the same colony and decrease the integrity estimator $M^+$ since the two ants did not accept each other but belong to the same colony, suggesting the integration of colony members is relatively low.
If two ants meet that do not belong to the same colony but accept each other then R5 is applied which will, over time, lead to the ability that smaller colonies - where many of these are initially formed in the beginning via R1 - getting integrated into the bigger colonies.
For all other cases, the default rule is applied and thus nothing happens.

One thing not explained until now is the acceptance function $Acceptance(ant_i, ant_j)$.
It will return true or false based on the comparison of the ants templates against their similarity and is defined by
\vspace{0.4em}
\begin{align*}
    Acceptance&(ant_i, ant_j) \leftrightarrow  \\
    &(Sim(ant_i, ant_j) > template_i) \land \\
    &(Sim(ant_i, ant_j) > template_j)
\end{align*}

Where $Sim(ant_i, ant_j)$ is the similarity, $0 \leq similarity \leq 1$.
A similarity of $1$ means that the two ants are completely similar, a similarity of $0$ means they are anti similar.
The similarity function needs to be defined based on the data set \textit{AntClust} is running on.
Remember that each tuple of the data set defines one ant by defining the genetics of that ant, letting each ant represent one data tuple inside the set.
If the data set is the simplistic one described above - just numbers between zero and one - a similarity function would simply be $ Sim_{1d}(x, y) = 1 - | x - y |$.
Therefore $Sim(ant_i, ant_j)$ would simply extract the genetics of the ants - which is then a single number for every ant representing one data point inside the data set - and put it into $Sim_{1d}()$ to compute the similarity.
In a more sophisticated data set, each tuple of the set might contain more than one feature and these features might not be just numbers but vectors.
For each feature, the user would need to define and tell \textit{AntClust} which similarity function should be used.
The mean similarity from all used similarity measures will thereafter be used to compute the total similarity of two ants.
\begin{align*}
    Sim(ant_i, ant_j) = \frac{1}{n_{sim}} \cdot \sum_w^{n_{sim}} Sim_w(x_w, y_w)
\end{align*}
Where $n_{sim}$ is the number of similarity measures or features, $Sim_w()$ one specific similarity function for the data type $w$, $x_w$ and $y_w$ the $w^{th}$ feature of the data tuple extracted from the ant genetic, where $x_w$ is extracted from $ant_i$ and $y_w$ from $ant_j$.

With the above similarity definition it is possible to run the algorithm on any data set if a similarity function can be specified by the user. 
\begin{figure*}
\centerline{\includegraphics[width=\textwidth]{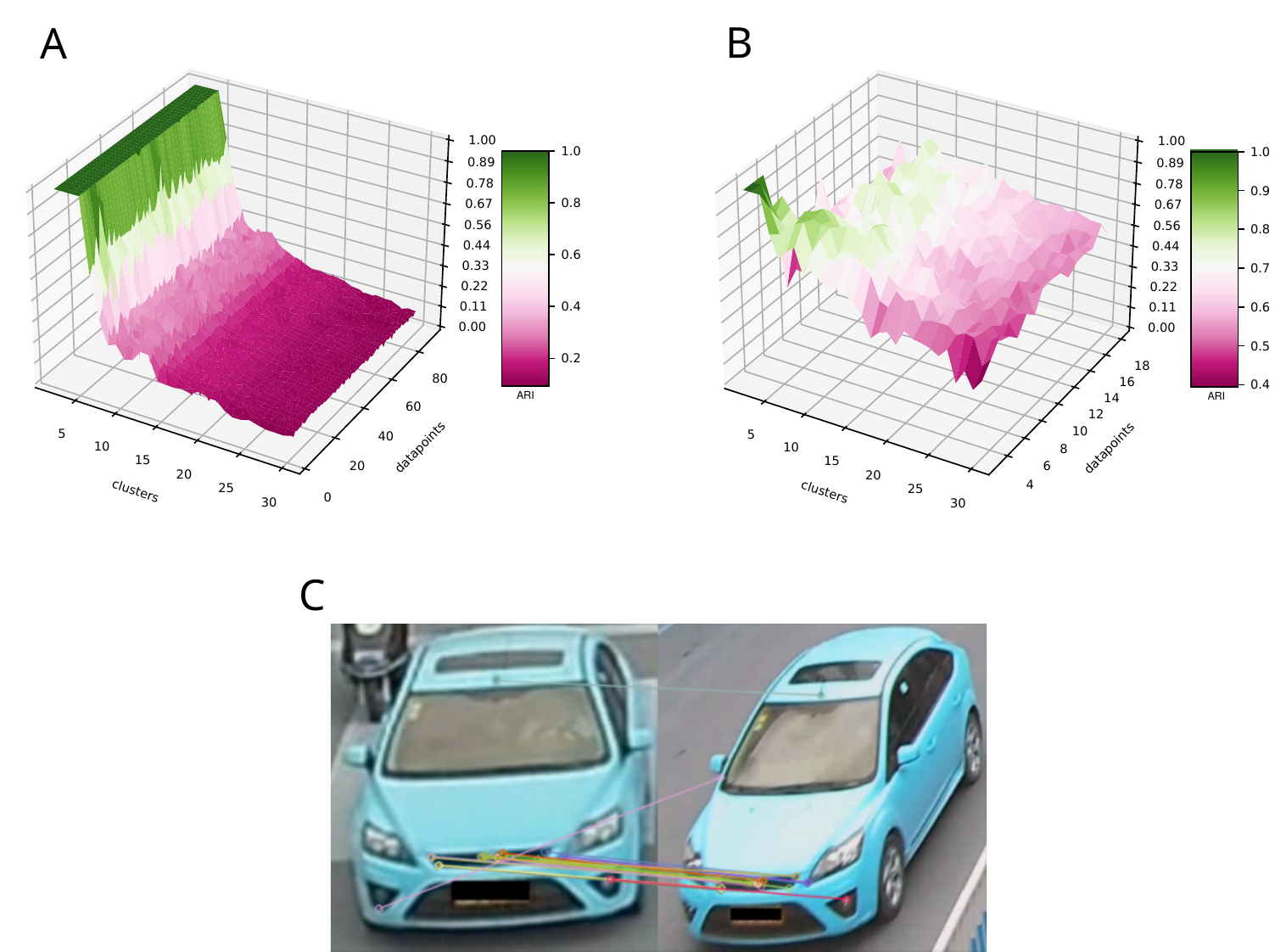}}
\caption{%
Adjusted Rand Index (ARI) score for the self generated float data set (\textbf{A}) and image data set (\textbf{B}).
The X-axis represent the number of clusters.
The Y-axis shows the amount of data tuples in each cluster.
This means that for each amount of clusters on the X- axis, all amounts of data tuples on the Y axis are tested in a clustering test.
The ARI score is shown on the Z-axis.
Here a score of $1$ indicates a perfect clustering - found labels match exactly the ground truth labels - where a score of $0$ indicates a random labeling of the data.
(\textbf{C}) shows the 20 best matching features between two images of the same car from the VeRi data set.%
}
\label{fig:cluster_performance_test}
\end{figure*}
%
%
% ----------------------------------
%  Proposed Method
% ----------------------------------
\section{Proposed Method}
\label{sec:proposed_method}
In most cases, clustering is performed on numerical data.
However, we apply clustering on images.\footnote{An image cluster library using traditional clustering methods is \textit{clustimage},\\ \url{https://github.com/erdogant/clustimage}}
The images we cluster are images of cars from the vehicle re-identification (VeRi) \footnote{https://vehiclereid.github.io/VeRi/} data set.
This set contains $50.000$ images of $776$ vehicles captured by $20$ different cameras.
A re-identification of a vehicle takes place if all images of one distinct vehicle are labeled into the same cluster.
This is the task for the clustering algorithm.
To achieve this task, a clustering algorithm uses a distance function or metric which tells how similar or anti-similar two data tuples are.
For numerical values, there exist many metrics and the most common used are the Manhattan or the Euclidean distance.
On images, distance metrics are not trivial to define.
The similarity of an image might not be defined by comparing the pixel values of each pixel and calculating the difference.
However, pictures usually contain certain points of interest, so-called features.
For example, an image of the sky might mostly contain blue parts, which are not very interesting when comparing the picture.
Yet an interesting part might be a bird flying through the sky.
To detect such features there exist many different feature detection methods.
One well-known feature detection method is the Harris Corner Detection algorithm which will detect the parts of the image where corners cross.
Harris corner detection has problems when the scale of the image changes.
There exit other more robust and faster methods like Scale-Invariant Feature Transform (SIFT) or Speeded-Up Robust Features (SURF).
However, these methods are patented.
A more performant and not patented method comes from the developers of the open source computer vision library \textit{OpenCV}.\footnote{https://opencv.org/}
Their method is called Oriented FAST and Rotated BRIEF (ORB), where BRIEF stands for Binary Robust Independent Elementary Features \cite{b4}.
The method is implemented in \textit{OpenCV} and is free to use for everyone.

We use the \textit{ORB} features to define a distance metric on images with it.
Features of images are image specific and can be compared to each other.
By comparing the features - which is usually done using the Manhattan norm - one retrieves a distance between features.
Thus it is possible to define a similarity metric for images with them.
\textit{OpenCV} contains feature matchers which allow to compare the features of the images.
A brute force matcher (BF) will compare each feature of the first image to all features of the second image by using a specified norm. 
There are more advanced methods for comparing image feature like the Fast Library for Approximate Nearest Neighbors (FLANN) matcher, which will utilize different algorithms to find the best matching descriptors.
We used the brute force matcher - as it fast on multi core CPU's - in combination with the hamming norm to find distances between the features.
These distances can then be used to evaluate the similarity between two pictures by comparing how many features from the first image can be found in the second image and how far the distance between the features is.
We only used the distance of the best matching feature.
This means that from each picture, only the distance from the best matching feature is used to define the distance metric of two images.
We know that this is problematic as if, for some reason, the feature exists in the two images but these are generally not similar, e.g. the same street sign is present in both images but the images show two completely different cars.
Using more features for distance calculation would need more tuning as distances between the features vary to a high degree and as such, it is not easy to normalize them correctly between zero and one - which is needed for \textit{AntClust}.
As such, by using more features there was a high mismatch in the similarity function, which did not always return $1$ if the images showing the same car.
The score was far below $1$ as scaling did not work properly.
This is a major problem and for gaining better results, it should be addressed as discussed in the final section \ref{sec:antclust_conclusion_future}.
%
%
% ----------------------------------
%  Experiments and Results
% ----------------------------------
\section{Experiments and Results}\label{sec:antclust_results}
We examine the clustering performance of \textit{AntClust} on an image data set.
We compare \textit{AntClust}\footnote{we tweak the default parameters to have more Ant template initialisation meetings by $\beta = 0.9$ and more Ant meetings in the meeting phase of the Algorithm, by setting $\alpha = 500$
} to other cluster algorithms such as DBSCAN, HDBSCAN and OPTICS\footnote{we use the scikit-learn implementation with the default parameters except for DBSCAN where we set eps=$0.33$ and min-samples=$2$} for reference.
Additionally, we perform testing on a self-defined artificial float data set to show that the clustering performance of \textit{AntClust} depends on the used ruleset.
% Once clustering of a data set took place the clustering result - the proposed labels for each data tuple - has to be evaluated.
To evaluate the clustering performance, we used the Adjusted Rand Index score.
Adjusted Rand Index is defined as a similarity measure that computes the differences in samples from a ground truth labeling to the provided labeling by the clustering algorithm.
The highest score is $1.0$, which indicates a perfect labeling of the data.
A score of $0.0$ indicates a random labeling.
If the score is negative, the labeling is even worse than a random labeling.

% Image dataset
We analyze the clustering performance for images in the following way.
From the VeRi data set, we select $30$ distinct cars, where for each car we take $18$ images showing the car from the front.
The images are used as a base to construct different clustering tasks.
We start with only taking the images of two cars to have a very simple clustering task of clustering the images into two different clusters.
Thereafter, images of three different cars, then four, until images of thirteen different cars are used to generate the task of finding three, four and finally thirteen clusters.
The clustering task is harder if more clusters need to be found.
To variate the tasks we change the amount of images in each cluster.
For each of the above tasks, we change the amount of images in each cluster from $3$ to $18$.
Results of these tasks show that the clustering performance worsens when more clusters are present - result can be seen in Figure \ref{fig:cluster_performance_test} (B).
For up to $10$ clusters, the performance stays relatively good.
Having more clusters in the clustering performance drops significantly.
It has no influence on how many images are inside each cluster.
The mean clustering performance only changes if the numbers of clusters increases, not the number of images inside each cluster.

% float assembly's
Declining in clustering performance for many clusters is expected due to the fact that there is only one rule in the Labroche rule set, which is creating new clusters - and this rule is only applied in the beginning of the clustering phases of \textit{AntClust}.
To test the hypothesis, we generated a simplistic, one-dimensional data set which contains only float numbers that belong to a cluster.
Each cluster has an integer as a pivot element.
The pivot elements are defined as $P = \{1,2,...,n\}$.
Each data point in a cluster differs from this pivot element by a range $R = [-0.1, 0.1]$.
The data set is then defined as $D = \{d_i | d_i = x + r, x \in P, r \in R\}$.
This creates a very clear and easy clustering task that can easily be extended to many clusters or data tuples in each cluster.
We generate $N$ clusters, each cluster having the exact same amount of data tuples $d$ in it.
We start with $N=2$ clusters, each with $d=3$ tuples.
Then increase the number of tuples up until $d=90$ in each cluster.
After that, we incremented the number of clusters by one and started again with $d=3$ tuples in each cluster, increasing them to $d=90$.
We proceeded until we reached $N=30$ clusters.
Meaning in the last test, we have $30$ clusters, each having $90$ data tuples in it.
The clustering result gets worse if more than $5$ clusters are present in the artificially generated data - as seen in Figure \ref{fig:cluster_performance_test} (A).
This supports the finding that \textit{AntClust} does not handle many clusters in a data set.
However, the tests revealed only the performance on this particular artificial data set.
Since \textit{AntClust} only sees the data in the form of similarity, this should however show the performance trend, the direction in which the performance for many clusters will evolve.

As reference, we test \textit{AntClust} against DBSCAN, HDBSCAN and OPTICS.
The task for all algorithms is to cluster data sets containing images of cars.
Each cluster contains $18$ images of cars.
The Number of cars - and thus the number of clusters to be found - is varied from $2$ to $30$.
For each algorithm, the found clustering partition is evaluated using the ARI score.
We pre-compute a distance matrix with our similarity function and provide it to the algorithms.
Therefore the clustering task is exactly the same for all algorithms.
Results show that it depends on the number of clusters which algorithm performs best.
In this task, \textit{AntClust} delivers the best clustering partitioning.
All ARI scores for the respective algorithms and their mean score are shown in Table \ref{tab:reference_test}.

% results end part 
\textit{AntClust} can be a clustering algorithm of choice if not many clusters are expected in the data.
A good reason for using \textit{AntClust} is that it does not need to know how many clusters reside in the given data set.
This is a significant improvement over clustering algorithms such as K-means which require the number of clusters to be found as an input parameter.

%
%
%
% results table
\begin{tiny}
\setlength{\tabcolsep}{6pt}
\begin{table}
\begin{center}
\caption{ARI clustering score for different amounts of clusters inside the car image data set. In the case of two clusters, images from two cars were used, and images of four cars were used to have four clusters.
Mean represents the mean ARI score for all cluster numbers.}
\label{tab:reference_test}
\resizebox{\textwidth}{!}{%
    \begin{tabular}{l | lllllllllllllll|l}
    \hline 
    %\hline\noalign{\smallskip}
    Clusters & 2 & 4 & 6 & 8 & 10 & 12 & 14 & 16 & 18 & 20 & 22 & 24 & 26 & 28 & 30 & Mean\\
    \hline
    AntClust & 0.31 & 0.74 & 0.66 & 0.78 & 0.66 & 0.69 & 0.69 & 0.66 & 0.63 & 0.65 & 0.59 & 0.61 & 0.6 & 0.57 & 0.53 & 0.62\\
    DBSCAN & 0.75 & 0.73 & 0.74 & 0.73 & 0.72 & 0.34 & 0.23 & 0.2 & 0.17 & 0.21 & 0.15 & 0.11 & 0.083 & 0.059 & 0.062 & 0.35\\
    HDBSCAN & 0.86 & 0.76 & 0.55 & 0.71 & 0.65 & 0.49 & 0.45 & 0.48 & 0.41 & 0.4 & 0.36 & 0.41 & 0.38 & 0.35 & 0.39 & 0.50\\
    OPTICS & 0.45 & 0.63 & 0.29 & 0.45 & 0.39 & 0.27 & 0.28 & 0.24 & 0.16 & 0.17 & 0.12 & 0.1 & 0.094 & 0.076 & 0.086 & 0.25\\ 
    \hline
    \end{tabular}
}
\end{center}
\end{table}
\setlength{\tabcolsep}{1.4pt}

\end{tiny}
%
%
%

%
% ----------------------------------
% Conclusion and Future Work
% ----------------------------------
\section{Conclusion and Future Work}\label{sec:antclust_conclusion_future}
We implemented \textit{AntClust} and used our implementation to cluster images of cars taken from the VeRi data set.
We used ORB features extracted from the images to define the similarity between images and matched the feature distances.
The results show that \textit{AntClust} performs well on our chosen data set, outperforming DBSCAN, HDBSCAN or OPTICS.
We found that clustering performance gets significantly worse for the artificially generated float assembly data set if many clusters are present. 
% This is to our understanding due to the used rule set which allows generating new clusters only in the beginning phase of the algorithm.

An improvement factor in future work is the similarity function.
We extract image features using ORB.
Currently, the similarity is calculated using only the distance between the nearest features in all obtained features, meaning only one feature is used to obtain the similarity.
As a future work, more features should be used to have a more robust match.
Additionally, the current approach is not color-aware.
It would be a benefit to include color-aware similarity measures.
Generally - as done in many classification tasks - using an ensemble of image similarity methods to improve the overall performance of the similarity function would be necessary to test.
We plan to test the performance of these new similarity functions on more than one image data set.
Having more than one image data set will give a better generalized picture of the performance of the \textit{AntClust} algorithm used on images.

\textit{AntClust} has the advantage that it does not need to know the numbers of clusters in the data beforehand - such as k-means.
Finding the number of clusters that best separates a given data set is not a trivial task.
Our tests show that despite not knowing how many clusters are existing in the data, still, a good clustering result can be obtained.
However, this is only true if the data does not contain to many clusters.
The reason for this is the following.
The rule set used by \textit{AntClust} should be able to generate and change clusters dynamically.
For example, the default Labroche rule set will create new clusters with its rule R1 and alters clusters with rule R5.
After the meeting phase, \textit{AntClust} will take care that only clusters with a high fitness exists by deleting clusters with lower fitness in the nest shrink step (see Table \ref{tab:antclust_steps}).
In theory, this should lead to the best partitioning, i.e. the "right" amount of clusters.
From the results obtained by Labroche et al., it can be seen that this works pretty well for data sets that have only up to four clusters \cite{b0}.
As we showed, if more clusters are inside the data, \textit{AntClust} struggles with generating enough clusters.
As suggested by Labroche et al., this might be due to the fact that the only rule that can create new clusters is rule R1.
The problem here is that R1 is only applied if, during a meeting, both ants do not have any labels.
As such, in the beginning, R1 is applied very often, whereas in the end, R1 will never be applied.
If a particular cluster is not formed in the beginning, or if it is deleted during runtime, it can not be recreated again.
Therefore it would be necessary to alter the rule set in a way that it allows for creating new clusters even in the later phases of the algorithm.
This is, however, a challenging task.
Our implementation opens the way for experimenting with different rule sets, makeing it easy to change and replace the rule set via an informal interface in \textit{Python}.

Future work will be on an automated rule set generation approach.
Neuronal networks could be trained and used as a rule set - and might be trained by neural evolution mechanisms such as NeuroEvolution of Augmenting Topologies (NEAT) \cite{b7}.
Rule sets or rule mechanisms could be evolved using genetic algorithms or a gene expression programming (GEP) approach.
This will result in a bio-inspired algorithm based on the chemical recognition system in ants and the evolutionary algorithm for more complete and accurate results.
It enables bio-inspired algorithms to enhance each other and become better together.

%---------------------------------------------------------
%               Bib
%---------------------------------------------------------


\begin{thebibliography}{1}
\bibitem{b0} Labroche, Nicolas \& E, Nicolas \& Venturini, Gilles. (2003). A New Clustering Algorithm Based on the Chemical Recognition System of Ants. 
%
\bibitem{b1} Labroche, Nicolas \& Monmarché, Nicolas \& Venturini, Gilles. (2003). AntClust: Ant Clustering and Web Usage Mining. 25-36. 10.1007/3-540-45105-6\_3. 
%
\bibitem{b2} B. Hölldobler and E.O. Wilson, The Ants, chapter Colony odor and kin recognition, 197–208, Springer Verlag, Berlin, Germany, 1990.
%
\bibitem{b3} D. Fresneau and C. Errard, ‘L’identité coloniale et sa reprsentation chez les fourmis’, Intellectica, 2, 91–115, (1994).
%
\bibitem{b4} Rublee, Ethan \& Rabaud, Vincent \& Konolige, Kurt \& Bradski, Gary. (2011). ORB: an efficient alternative to SIFT or SURF. Proceedings of the IEEE International Conference on Computer Vision. 2564-2571. 10.1109/ICCV.2011.6126544. 
%
\bibitem{b5} Implementation of the Algorithm \\ \url{https://gitlab.com/Winnus/antclust}
%
\bibitem{b6} Xinchen Liu, Wu Liu, Tao Mei, Huadong Ma: PROVID: Progressive and Multimodal Vehicle Reidentification for Large-Scale Urban Surveillance. IEEE Trans. Multimedia 20(3): 645-658 (2018)
%
\bibitem{b7} Kenneth O. Stanley, Risto Miikkulainen; Evolving Neural Networks through Augmenting Topologies. Evol Comput 2002; 10 (2): 99–127. doi: \url{https://doi.org/10.1162/106365602320169811}
%
\bibitem{b8} AntClust Python Implementation Winfried Oed 2022 \url{https://gitlab.com/Winnus/antclust}

% \bibitem{b100} Figure \ref{fig:feeling_wheel} \url{https://commons.wikimedia.org/wiki/File:The_Feeling_Wheel.png}

\end{thebibliography}
\end{document}